\documentclass[10pt,preprint]{aastex}
\usepackage{graphicx}
\usepackage{epstopdf}
\shorttitle{Star Formation in the Horsehead Nebula}
\begin{document}
\title{An Infrared Census of Star Formation in the Horsehead Nebula}
\author{Brendan P. Bowler$^1$\footnotetext[1]{Institute for Astronomy, University of Hawai'i, 2680 Woodlawn Drive, Honolulu, HI 96822; bpbowler@ifa.hawaii.edu}, William H. Waller$^2$\footnotetext[2]{Department of Physics and Astronomy, Tufts University, Medford, MA 02155; william.waller@tufts.edu}, S. Thomas Megeath$^3$\footnotetext[3]{Department of Physics and Astronomy, University of Toledo, Toledo, OH 43606; megeath@physics.utoledo.edu}, Brian M. Patten$^{4,5}$\footnotetext[4]{National Science Foundation, 4201 Wilson Blvd, Arlington, VA 22230; bpatten@nsf.gov}\footnotetext[5]{Harvard-Smithsonian Center for Astrophysics, 60 Garden St., Cambridge, MA 02138}, and Motohide Tamura$^6$\footnotetext[6]{National Astronomical Observatory of Japan, Osawa, Mitaka, Tokyo 181-8588, Japan; motohide.tamura@nao.ac.jp} }

\begin{abstract}

At $\sim$ 400 pc, the Horsehead Nebula (B33) is the closest radiatively-sculpted pillar to the Sun, but the state and extent of star formation in this structure is not well understood.  We present deep near-infrared (IRSF/SIRIUS $JHK_\mathrm{S}$) and mid-infrared (\emph{Spitzer}/IRAC) observations of the Horsehead Nebula in order to characterize the star forming properties of this region and to assess the likelihood of triggered star formation.  Infrared color-color and color-magnitude diagrams are used to identify young stars based on infrared excess emission and positions to the right of the Zero-Age Main Sequence, respectively.  Of the 45 sources detected at both near- and mid-infrared wavelengths, three bona fide and five candidate young stars are identified in this 7$'$$\times$7$'$ region.  Two bona fide young stars have flat infrared SEDs and are located at the western irradiated tip of the pillar.  The spatial coincidence of the protostars at the leading edge of this elephant trunk is consistent with the Radiation-Driven Implosion (RDI) model of triggered star formation.  There is no evidence, however, for sequential star formation within the immediate $\sim$ 1$\farcm$5 (0.17 pc) region from the cloud/H \small II \normalsize region interface.  
\end{abstract}

\section{Introduction}

Star formation is a multi-faceted phenomenon that can occur spontaneously or as a reaction to external influences.  A scenario that has gained acceptance over the last decade to account for many of the small-scale properties of star-forming regions involves the direct (winds and radiation) and indirect (expansion of H \small II \normalsize region) triggering of star formation by massive young stars.  In each case, triggered star formation is the result of an externally-induced compression of a molecular cloud.  It has been invoked to explain the apparent enhanced luminosities of protostars near H \small II \normalsize regions (\citealt{Motoyama:2007p488}; \citealt{Sugitani:1989p457}), the age spread of young clusters (\citealt{Lee:2007p520}; \citealt{Lee:2005p442}), synchronized star formation at the peripheries of young clusters (\citealt{Smith:2007p523}), small-scale sequential star formation (e.g., \citealt{Ikeda:2008p521}; \citealt{Getman:2007p480}; \citealt{Matsuyanagi:2006p462}; \citealt{Sugitani:1995p476}), and star formation in bright-rimmed clouds (e.g., \citealt{Sugitani:1994p482}; \citealt{Sugitani:1991p449}).

There are two leading models through which triggered star formation is plausible near H \small II \normalsize regions.  \citet{Chen:2007p1222} outline these cases and describe likely examples of this phenomenon.  In the ``collect and collapse'' model (\citealt{Elmegreen:1977p474}; \citealt{Hosokawa:2006p437}), an expanding H \small II \normalsize region sweeps up material into a dense layer bordering the H \small II \normalsize region and the parental molecular cloud.  This shell of compressed gas and dust then fragments and gravitationally collapses to form new stars.  In Radiation-Driven Implosion (RDI; \citealt{Bertoldi:1989p453}), the photoevaporation of a bright-rimmed cloud (BRC) by the intense UV radiation of nearby massive young stars creates an ionization shock-front at the surface of the cloud.  The inward pressure initiates the formation of new cores, or the compression of pre-existing ones, that then collapse to form a second generation of stars.

Evidence for triggered star formation usually relies on the spatial distribution and gradient in evolutionary phases of young stars near a BRC (RDI model) or the observation of a dense, fragmented shell of gas with newly formed stars surrounding an H \small II \normalsize region (collect and collapse model).  In the case of RDI, several authors have discussed the difficulties of determining whether young stars are actually induced to form or whether they are merely being exposed by photoevaporation (e.g., \citealt{Dale:2007p481}).  The real picture is probably some combination of these scenarios, where the compression of preexisting clumps may produce an acceleration of what would have been local spontaneous star formation.  However, even before the collapse of a core to form a protostar, the influence of massive stars on dense cores has been observed in the large velocity widths of cores near the M42 H \small II \normalsize region (\citealt{Ikeda:2007p444}), leaving no doubt that star formation near massive stars is not an isolated process.

Pillars (or elephant trunks) are commonly observed in star-forming regions and, although differing in shape and size, are thought to have similar formation mechanisms.  As a molecular cloud is photoionized by a massive young star, a pre-existing dense clump will photoevaporate at a slower rate than the surrounding more tenuous material.  A pillar will form when such a clump effectively protects the region behind it from photoevaporation.  Although they are still being eroded away, they do so at slower rates than the cloud material from which they protrude.   In this scenario these pillars are the astronomical analogues of hoodoo formations in geology.  Other theories of pillar formation include a variety of instability models (e.g., \citealt{Spitzer:1954p524}; \citealt{Frieman:1954p525}) and twisted magnetic flux ropes   (\citealt{Carlqvist:2003p204}).  The combination of stellar winds and an expanding ionization front probably influences the formation of these structures as well (\citealt{Schneps:1980p527}).  These pillars point radially in the direction of the weathering source and are often characterized by bright rims in H$\alpha$ emission.   Several studies have employed Smoothed Particle Hydrodynamical models in an attempt to better-understand pillar formation and evolution.  In these models core formation and eventual core collapse occurred at the leading edges of pillars as a direct result of the pressure caused by photoevaporation at the cloud's surface (\citealt{Gritschneder:2007p1226}; \citealt{Miao:2006p528}).   Young stars at the tips of such pillars have been observed in over a dozen cases, including the Elephant Trunk Nebula (\citealt{Reach:2004p447}), the Eagle Nebula (\citealt{Linsky:2007p198}; \citealt{Sugitani:2002p217}; \citealt{Fukuda:2002p202}), the Cone Nebula (\citealt{Dahm:2005p503}), several pillars in 30 Doradus (\citealt{Walborn:2002p529}), and many other, lesser-known pillars  (\citealt{Sugitani:1995p476}; \citealt{Carlqvist:2003p204}; \citealt{Ubeda:2007p281}; \citealt{Wang:2007p530}).

At a distance of $\sim$ 400 pc (\citealt{AnthonyTwarog:1982p531}), the Horsehead Nebula (or B33: \citealt{Barnard:1919p532}) is the closest elephant trunk to the Sun and represents an ideal laboratory to study the interaction between massive stars and molecular clouds.  It extends westward from the southern part of L1630 into the H \small II \normalsize region IC 434 and points radially toward the OB system $\sigma$ Orionis, itself at the center of the $\sim$ 3 Myr $\sigma$ Orionis cluster (\citealt{Hernandez:2007p387}; \citealt{Oliveira:2006p250}).  The Horsehead has a length of roughly 0.8 pc and it is located about 4 pc in projection from the source of its photoerosion, $\sigma$ Ori A (SpT: O9.5V).  This is likely a gross underestimation of their actual separation though; a wide range of distances to the $\sigma$ Ori cluster have been published but typically fall between 350 pc and 470 pc (\citealt{Perryman:1997p534}; \citealt{Hernandez:2005p536}, \citealt{deZeeuw:1999p538}), with the most recent main-sequence fitting placing this cluster at 420 $\pm$ 30 pc (\citealt{Sherry:2008p539}).  The accurate distance to both the Horsehead and $\sigma$ Ori is important because the potential to trigger star formation depends strongly upon the separation between the cloud and the massive star(s) (\citealt{Megeath:2002p448}).  An historical account of the discovery of the Horsehead Nebula and subsequent early plate recordings is given in \citet{Pound:2003p235}, and a comprehensive account of more recent work is presented in Meyer et al. (in press).  

The physical characteristics of the Horsehead have been studied by many authors.  Its molecular mass is roughly 27 M$_{\odot}$ (\citealt{Pound:2003p235}; \citealt{Philipp:2006p227}; \citealt{Lada:1991p240}); the gas density varies between 3-7 $\times$ $10^4$ cm$^{-3}$ (\citealt{Philipp:2006p227}) in most of the Horsehead and reaches a peak of 6 $\times$ $10^5$ cm$^{-3}$ in the denser of two sub-mm cores (B33-SMM1, \citealt{WardThompson:2006p540}).  Its gas temperature ranges from 10 K to roughly 20 K (\citealt{Pety:2007p541}; \citealt{Philipp:2006p227}).  The formation and evolution of B33 and its characteristic substructure have been debated in the literature, but little consensus has been reached (\citealt{Reipurth:1984p544}; \citealt{WarrenSmith:1985p207}; \citealt{Neckel:1985p545}; \citealt{HilyBlant:2005p546}).

The photodissociative properties of this region have been studied in depth (e.g., \citealt{Compiegne:2007p374}; \citealt{Pety:2005p542}; \citealt{Teyssier:2004p548}; \citealt{Abergel:2003p516}), but few authors have focused on its star-forming properties.  The first targeted search for embedded young stars in the Horsehead was performed by \citet{Reipurth:1984p544}.  They presented shallow near-infrared images which revealed many optically-invisible infrared sources, one of which had a large near-infrared excess (B33-1).  This same source was detected by IRAS (IRAS 05383--0228).  Two other IRAS sources  were also found (IRAS 05384--0229, IRAS 05386--0229), and it has been suggested that they might be embedded young stars (\citealt{Pound:2003p235}).  ASCA observations uncovered two X-ray sources in the region (\citealt{Nakano:1999p502}; \citealt{Yamauchi:2000p225}) but their status as young stars remains unclear.  In this study we present near-infrared (NIR) and mid-infrared (MIR) imaging observations of the Horsehead Nebula in order to clarify its overall state of star formation and to assess the likelihood of external triggering activity.  

This paper is organized in the following manner.  In $\S$ 2 we introduce the IRSF and \emph{Spitzer}/IRAC observations, including data reduction, calibration, and analysis.  The results of the NIR and MIR color-color diagrams and the NIR color-magnitude diagrams are described in $\S$ 3.  In $\S$ 4 we discuss the current state of star formation within this pillar.  We conclude by summarizing our results in $\S$ 5.

\section{Observations and Data Reduction}

\subsection{Near-Infrared Observations}

\begin{figure}
 \resizebox{\textwidth}{!}{\includegraphics{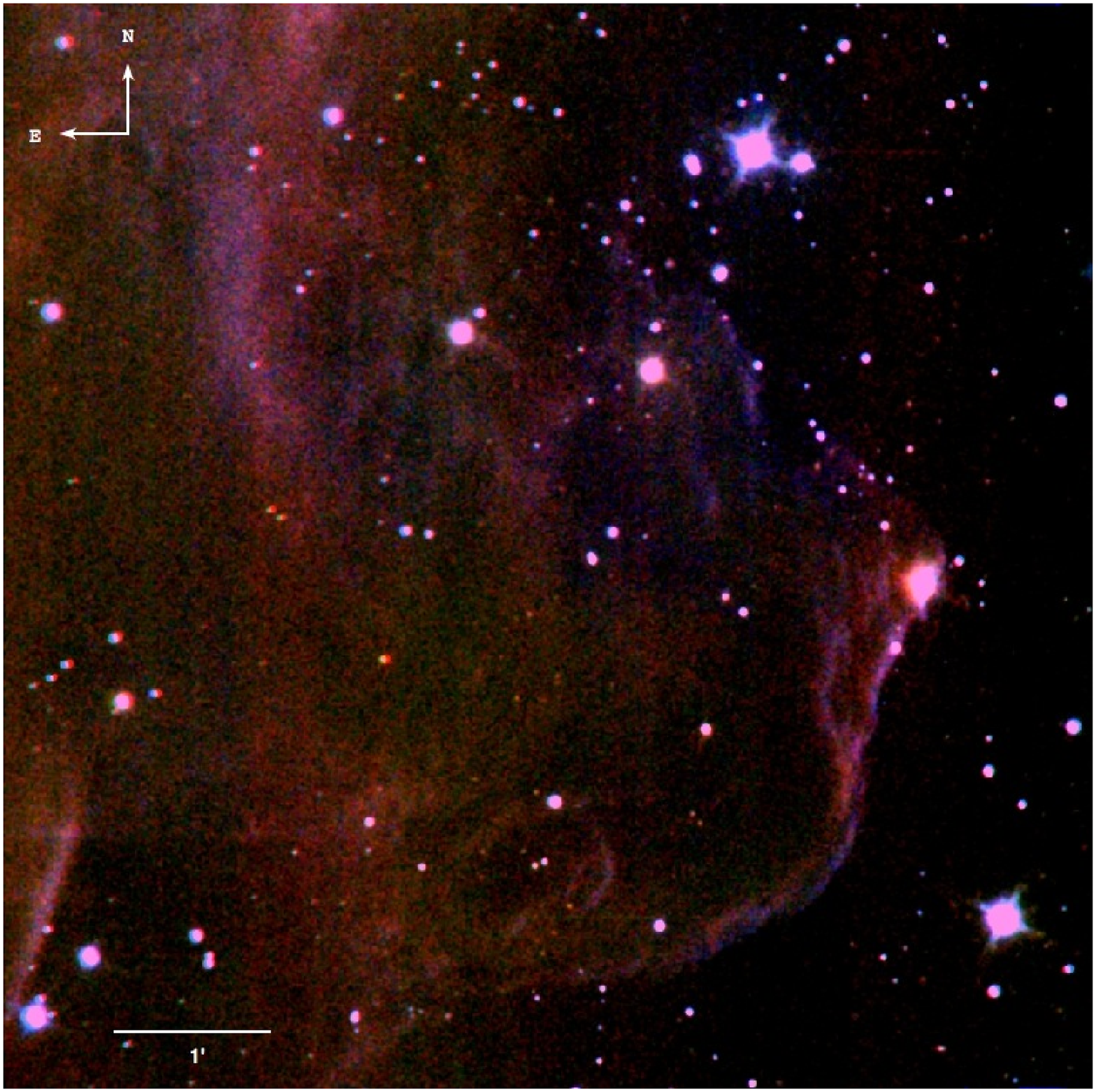}}
\caption{Three-color composite near-infrared image of the Horsehead Nebula.  Blue, green, and red represent the $J$-, $H$-, and $K_S$-band IRSF/SIRIUS images. The images have been Gaussian smoothed for better rendering.  \label{irimage}}
\end{figure}

Near-infrared observations were obtained on 13 March 2002 with the Simultaneous-3 color InfraRed Imager for Unbiased Survey (SIRIUS) (\citealt{Nagashima:1999p246}; \citealt{Nagayama:2003p1244}) mounted on the 1.4 m InfraRed Survey Facility (IRSF) telescope, which is part of the South African Astronomical Observatory (SAAO).  The SIRIUS camera allows for efficient simultaneous multi-band imaging.  Focused light is split by two dichroic mirrors to the three $J$ ($\lambda $ = 1.25 $\mu$m), $H$ ($\lambda$ = 1.63 $\mu$m), and $K_S$ ($\lambda$ = 2.14 $\mu$m) near-infrared filters.  The detectors are 1024 $\times$1024 HgCdTe (HAWAII) infrared arrays with pixel scales of 0$\farcs$45 pix$^{-1}$, producing a 7$'$$\times$7$'$ field of view.  Observations were taken in a dithered pattern to account for bad pixels and cosmic rays.  Three sets of ten 30 s images were obtained in each filter, producing an effective integration time of 900 s for each coadded image.  Sky frames were centered at $\alpha$$_{J2000.0}$ = 05:41:41, $\delta$$_{J2000.0}$ = $-$02:26:30 and twilight flat fields were obtained at dusk.  Dark current subtraction, sky subtraction, and flat field division were performed on site using the pipeline reduction procedure outlined in \citet{Nagashima:1999p246}.  The reduced frames were then shifted, trimmed and median-combined using the IMALIGN  and IMCOMBINE  tasks in the NOAO/IMAGES package from the Image Reduction and Analysis Facility (IRAF)$^7$\footnotetext[7]{IRAF is distributed by the National Optical Astronomy Observatories, which are operated by the Association of Universities for Research in Astronomy, Inc., under cooperative agreement with the National Science Foundation.}.  The combined $J$-, $H$-, and $K_S$-band NIR image of the Horsehead is shown in  Figure \ref{irimage}.

The source detection algorithm DAOFIND in the NOAO/DIGIPHOT/APPHOT package in IRAF was run separately for each of the three coadded images.  A visual inspection of each coadded frame was performed to ensure that no sources were missed and that obvious contaminants were excluded.  Typical Full-Width at Half-Maximum (FWHM) values for the three bands were 1$\farcs$2 and remained constant throughout the observations.  Aperture photometry was performed using the APPHOT task in IRAF using an aperture radius of 1$\farcs$8 with inner and outer sky annulus radii of 4$\farcs$5 and 7$\farcs$2, respectively.  These values were chosen based on a series of tests using different aperture radii to optimize the signal to noise ratio of the final flux.  A 3 $\sigma$ cutoff above the median value across the entire image was employed for each sky annulus to avoid contamination from nearby stars.  
Two bright sources were saturated in all three bands and one source corresponding to a known candidate young star was saturated in the $K_S$ band (B33-1).  2MASS magnitudes for all three bands were used for these three saturated stars.

The initial criterion  for source determination was a 5 $\sigma$ detection in at least two bands.  Later a second criterion that final, 2MASS-calibrated \emph{J}$-$\emph{H} and \emph{H}$-$\emph{K$_S$} color errors be less than 0.15 mag was employed.  The former criterion ensured that detections over multiple bands were indeed real and was used for initial screening.  The latter criterion was used after color transformations to the 2MASS photometric system and associated errors were determined.  It was then applied to all sources so that the \emph{J}$-$\emph{H} and \emph{H}$-$\emph{K$_S$} colors would have reasonably small errors for placement on the NIR color-color diagram (see $\S$2.1.1).  A total of 119 NIR sources satisfied both criteria.

Limiting magnitudes are roughly $J$=19.0, $H$=18.0, and $K_S$=17.5.  As a reference, at the distance of B33 (roughly 400 pc), a 1 Myr brown dwarf at 0.08\emph{M$_\odot$} has a $K_S$-band magnitude of $\sim$13 (based on the evolutionary tracks of Barrafe et al. 1998).  This survey is therefore within reach of slightly  embedded low-mass young stars and much more heavily embedded solar-mass young stars.

\subsubsection{Calibration Using 2MASS Photometry}
The IRSF sources were calibrated using the 2MASS catalogue (\citealt{Skrutskie:2006p589}).  The criteria for selecting 2MASS point sources within the IRSF field of view included the best photometric quality flags (labeled ``A'' in the 2MASS Point Source Catalogue, with a signal to noise ratio (SNR) $\geq$ 10) in at least two bands and a flag of at least second-best photometric quality in the third band (labeled ``B,'' with a SNR $\geq$ 7).  An additional criterion that sources have magnitudes between roughly 12 and 15 in the \emph{K$_{S}$} band was imposed as 2MASS is most sensitive over these magnitudes.  A total of 23 bright 2MASS sources were used to calibrate the IRSF magnitudes.

The basic transformation equations between filter systems were used to transform colors and magnitudes from the measured IRSF instrumental magnitudes to the standard 2MASS filter system (see, e.g., \citealt{Carpenter:2001p159}).  The coefficients and zero-point offsets were determined by applying a linear weighted least-squares fit to the appropriate color and magnitude relationships.  As expected, there was very little color dependence between the two systems.  The transformation equations were applied to all instrumental magnitudes.  SIRIUS does not record accurate sky coordinates and so all source coordinates were derived using the astrometry calculator \emph{Astrom}$^8$\footnotetext[8]{Distributed through the Starlink Software Collection, which is currently run by the Joint Astronomy Centre of the UK Science and Technology Facilities Council} with the 23 brightest 2MASS stars as reference positions for the interpolation of coordinates.

\subsection{Mid-Infrared Observations}

\begin{figure}
 \resizebox{\textwidth}{!}{\includegraphics{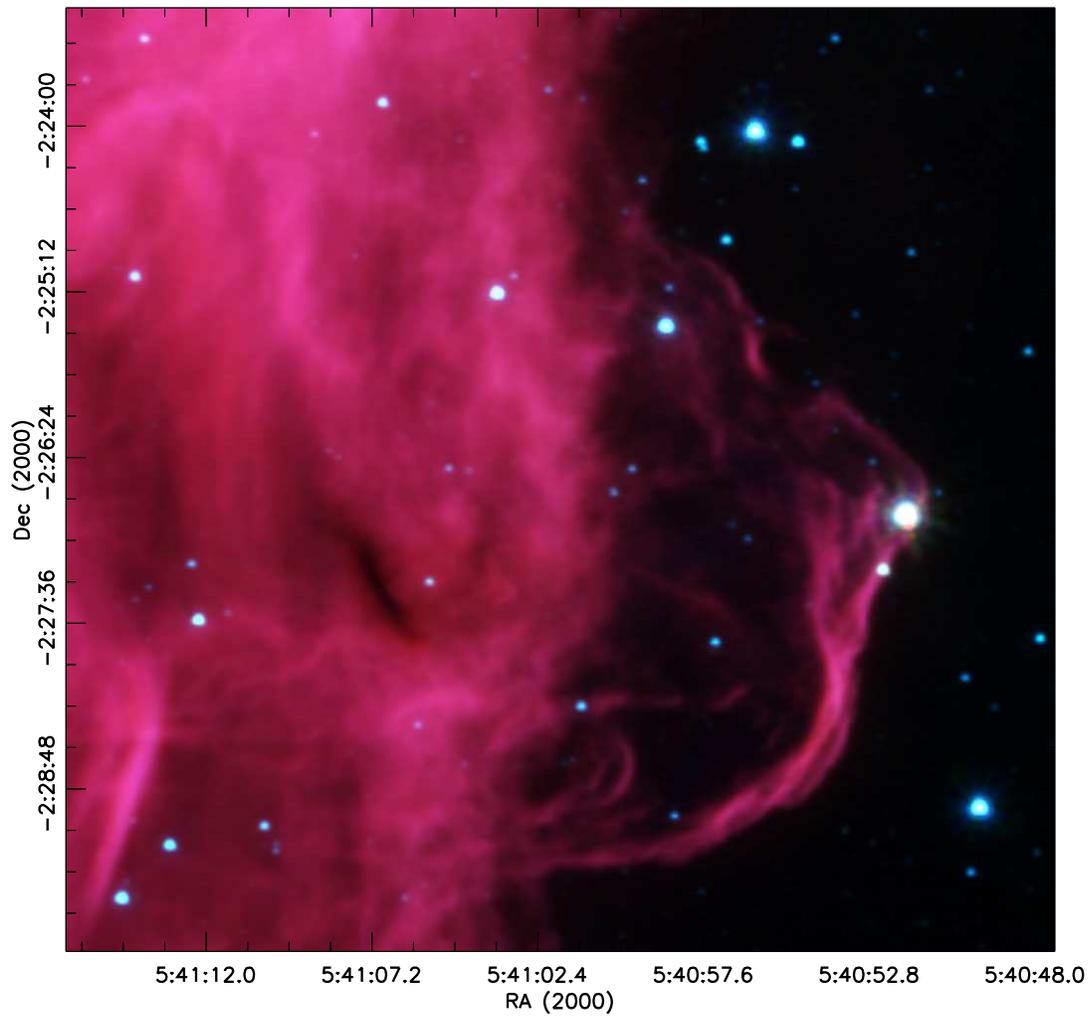}}
\caption{\emph{Spitzer} IRAC image of the Horsehead Nebula.  The image is a three color composite constructed from
the 3.6 $\mu$m (blue), 4.5 $\mu$m (green) and 8.0 $\mu$m (red) images.  The extensive nebular emission
is due to UV heated hydrocarbons.  \label{hhspitzer}}
\end{figure}

The Infrared Array Camera (IRAC) on board \emph{Spitzer Space Telescope} is a four-channel camera that obtains simultaneous broad-band images in 3.6, 4.5, 5.8, and 8.0 $\mu$m (\citealt{Fazio:2004p342}).  \emph{Spitzer}/IRAC observations of the Horsehead were obtained as part of the program titled \emph{An IRAC Survey of the L1630 and L1641 (Orion) Molecular Clouds} (PI: S.T. Megeath; PID43).  Two epochs of observations separated by about six months were used to construct two 12 s ``high dynamic range frames.''  These frames consist of a 12 s frame with 10.4 s of integration time and a 0.6 s frame with 0.4 s of integration time.  The total integration time is 41.6 s for the long frames and 1.6 s for the short frames.  The IRAC color-composite image of the Horsehead is shown in Figure \ref{hhspitzer}.

Photometry of point sources was initially obtained from the Basic Calibrated Data (BCD) from the Spitzer Science Center S14 data pipeline.  The BCD data were mosaiced using the WCSmosaic IDL package developed by Robert Gutermuth.  This software rejects cosmic rays by identifying deviant pixel values in overlapping frames, applies a frame-by-frame distortion correction, derotation, and subpixel offsetting in a single transformation, and matches the background by comparing signal levels in overlapping images.

The photometry was extracted using PHOTVIS, an IDL photometry and visualization tool developed by Robert Gutermuth.  It performs a spatial filtering of the data, computes a noise map from the filtered data, and then searches for point sources that exceed a source detection threshold.  Aperture photometry was obtained for each of the sources using an aperture of 2 pixels and a sky annulus of 2-6 pixels.  Each pixel is 1$\farcs$2$\times$ 1$\farcs$2.  The data were then bandmerged and sources were considered coincident when they were within 1$\farcs$0.  If two sources satisfied that criterion then the closest sources were then identified as coincident.  The zero points used to obtain the magnitudes (in DN/S per BCD image) were 19.6642, 18.9276, 16.8468, and 17.3909, for the 3.6, 4.5, 5.8, and 8.0 $\mu$m bands, respectively.  The limiting magnitudes are [3.6] $\simeq$ 17.0, [4.5] $\simeq$ 16.6, [5.8] $\simeq$ 15.8, and [8.0] $\simeq$ 13.5 for errors $\leq$ 0.1 mag in the first two IRAC channels and $\leq$ 0.2 mag in the last two channels.  A total of 61 sources were observed with IRAC in at least one band.

\subsubsection{Rejection of Extragalactic Contaminants}

Galaxies with high star formation rates occupy the same region of the NIR color-color diagram as do young stars (\citealt{Geller:2006p590}).  Galaxies and young stars also occupy a similar region of the [3.6]--[4.5] vs [5.8]--[8.0] color-color diagram (\citealt{Gutermuth:2008p492}).  It is therefore imperative to remove these objects from the sample.  \citet{Hernandez:2007p387} describe two methods for rejecting extragalactic objects from IRAC data; we adopt both of these procedures in this study.  Roughly half of the sources with [3.6] $>$ 14.5 are extragalactic in nature (\citealt{Fazio:2004p591}), so we reject sources above that limit.  \citet{Gutermuth:2008p492} identify regions of the [3.6]$-$[5.8] vs. [4.5]$-$[8.0] and the [4.5]$-$[5.8] vs. [5.8]$-$[8.0] color-color diagrams that are dominated by polycyclic aromatic hydrocarbon-rich galaxies with active star formation.  We eliminated probable contaminants by identifying sources in those regions and removing them from the sample.  

\subsubsection{Merging the NIR and MIR bands}

Finally, the NIR and MIR sources were bandmerged by selecting for IRAC sources within 3\farcs6 (= 3$\times$FWHM$_{J-band}$) of each IRSF point source.  If more than one IRAC source satisfied that requirement then the nearest one was selected.  The final sample used in this study consisted of 45 sources with $J$, $H$, $K_S$ magnitudes and a detection in at least one IRAC band.

\section{Results}

\subsection{Identification of Young Stellar Objects}

A source is considered a candidate young stellar object (YSO) if there is evidence for an excess of IR emission that cannot be accounted for by photospheric emission alone, indicative of  a warm inner circumstellar disk.  Candidate YSOs are identified using the following criteria: sources must have an IR excess in \emph{at least two of the six}  NIR, MIR, and combination of NIR and MIR  color-color diagrams.  Alternatively, if the source is detected in all four IRAC channels, it is considered a candidate YSO if it has an excess in the MIR color-color diagram and has a spectral index $>$ --2.56, which allows for the inclusion of transition or anemic disks whose spectral energy distributions only break from the photospheric emission near 8 $\mu$m.  Furthermore, the positions of candidate YSOs in the NIR color-magnitude diagrams must be consistent with youth at the distance of the Horsehead (to the right of the Zero-Age Main Sequence).  If a source exhibits an excess in a shorter-wavelength color-color diagram then it must also have an excess in the longer-wavelength IR color-color plots (depending on which IRAC bands it was detected in) to remain as a candidate YSO, as young stars with warm inner disks will have an excess at both NIR and MIR wavelengths.  However, the converse isn't required: if the inner disk is cleared then the excess emission may only begin to appear at MIR wavelengths.  The 3.6 to 8.0 $\mu$m spectral indices, the spectral energy distributions, and the positions in color-color diagrams are used together to distinguish bona fide YSOs from candidate YSOs and to assess infrared evolutionary classes.

\subsection{Infrared Color-Color Diagrams}

\subsubsection{Near-Infrared Color-Color Diagram}

\begin{figure}
  \resizebox{\textwidth}{!}{\includegraphics{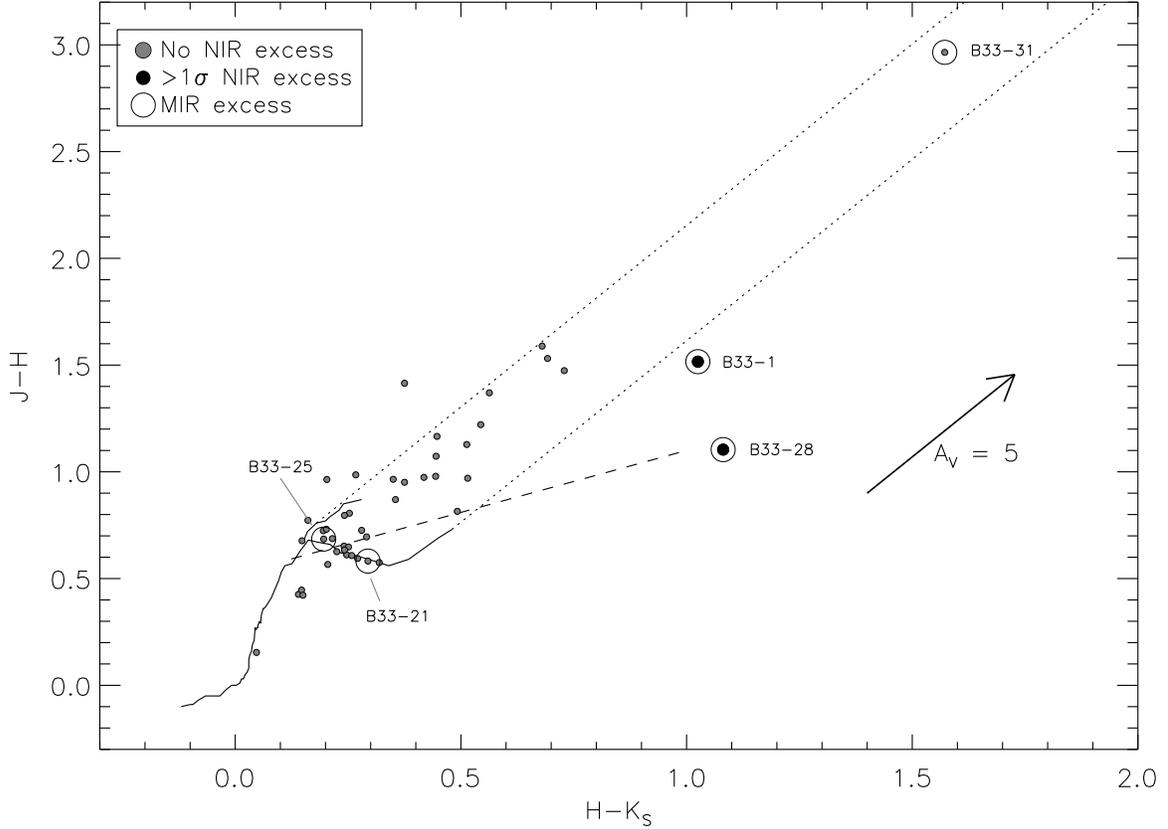}}
  \caption{$J-H$ vs. $H-K_S$ NIR color-color diagram.  Sources $\geq$ 1 $\sigma$ to the right of the right-most reddening envelope (\emph{dotted lines}) have an excess of NIR emission, likely caused by a warm circumstellar disk.  They are plotted as \emph{filled black circles}.  \emph{Solid lines} represent the empirical locus of main sequence stars and giants as determined by Bessell $\&$ Brett (1988).  An analagous T Tauri Star locus was found by Meyer, Calvet, $\&$ Hillenbrand (1997) and is plotted as a \emph{dashed line}.  MIR excess sources from the 4-channel IRAC color-color digram (Figure \ref{midircc}, top) are overplotted as \emph{open circles} and are labeled. All sources without a NIR excess are plotted as \emph{filled gray circles}.  The reddening vector is derived from \citet{Tokunaga:2000p1685}. \label{jhkcc} } 
\end{figure}

The NIR color-color diagram has long been recognized as an important tool in the identification of young stars with warm circumstellar disks (\citealt{Lada:1992p156}).  \citet{Bessell:1988p162} demonstrated that main sequence stars and giants on the \emph{J}$-$\emph{H} vs. \emph{H}$-$\emph{K$_S$} diagram sit on characteristic loci.  A similar empirical locus of Classical T Tauri stars was later found by \citet{Meyer:1997p592}.  Interstellar reddening will affect the position of a star on this diagram by shifting it to redder $J-H$ and $H-K_S$ colors, as is indicated by the arrow for $A_V$ = 5 mag in Figure \ref{jhkcc}.  Sources redward of the right-most reddening envelope are good candidates for young stars with warm disks, although young stars with and without disks may still reside in the ``normal'' portion of this plot.  We select sources in Figure \ref{jhkcc} with $H-K_S$ $\geq$ 1 $\sigma$ from the right-most reddening envelope to be candidate young stars.  There are two NIR excess sources (B33-1 and B33-28), one of which was detected by 2MASS (B33-1).  MIR excess sources in the [3.6]$-$[4.5] vs. [5.8]$-$[8.0] color-color diagram (see $\S$ 3.2.2 and Figure \ref{midircc}, below) are overplotted as open circles.

\subsubsection{Mid-Infrared Color-Color Diagrams}

MIR color-color diagrams are useful tools to assess the IR class of young stars and can help distinguish between highly reddened normal stars and those with true IR excesses caused by emission from cirumstellar disks.  \citet{Allen:2004p1016} plotted the positions of Class I (accreting envelopes and disks) and Class II (accreting disks) YSO models in the [3.6]$-$[4.5] vs. [5.8]$-$[8.0] diagram over a wide range of parameters, including accretion rates, disk sizes, and inclinations.  Unique regions were identified as being the loci of Class I and Class II objects.  The positions of young stars in that diagram  were subsequently shown to be in good agreement with these models (\citealt{Megeath:2004p294, Megeath:2005p1043}; \citealt{Hartmann:2005p433};  \citealt{Luhman:2006p1088}; \citealt{Balog:2007p1101}; \citealt{Winston:2007p489}).  Young stars with anemic disks or no disks at all (Weak-lined T Tauri stars/Class III IR sources) have similar NIR and MIR colors to main sequence dwarfs, where photospheric emission dominates over any slight disk emission (\citealt{Hartmann:2005p433}; \citealt{Hernandez:2007p387}); young stars with highly evolved disks are therefore missed in this type of survey.

\begin{figure}
  \resizebox{\textwidth}{!}{\includegraphics{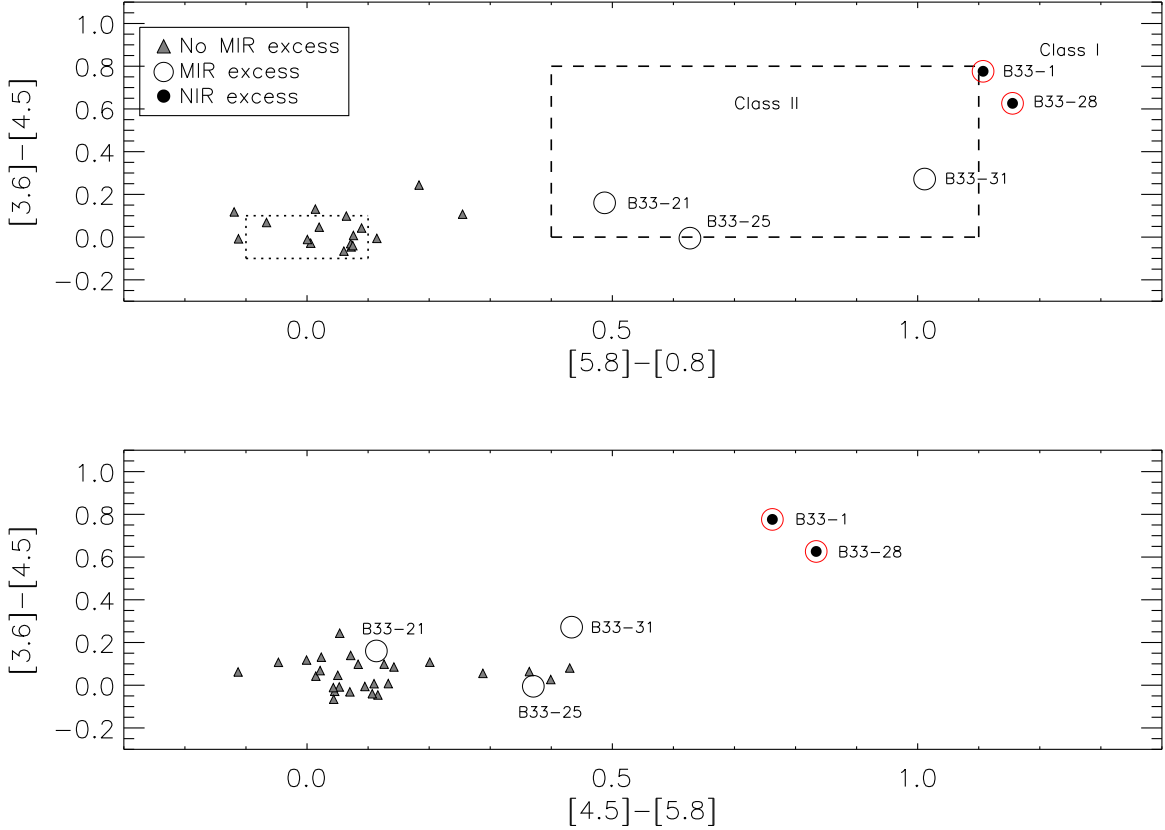}}
  \caption{[3.6]--[4.5] vs. [5.8]--[8.0] color-color diagram for 22 sources detected in all four IRAC channels (\emph{top}).  The five sources that exhibit MIR excesses are labeled (\emph{open circles}). Three fall in the Class II region and two fall in the Class I region (from Allen et al. 2004, Hartmann et al. 2005, and Luhman et al. 2006).  The two NIR excess sources from Figure \ref{jhkcc} are plotted as \emph{filled black circles}.  Those with no MIR excess are plotted as \emph{filled gray triangles}.  Sources detected in the first three channels of IRAC are plotted in the [3.6]--[4.5] vs. [4.5]--[8.0] diagram (\emph{bottom}).  The spread between IR evolutionary classes is smaller than that for the 4-channel color-color diagram, so this plot is not used to select YSO candidates; it is nevertheless instructive to show the similar trend of a red [4.5]--[5.8] color for candidate YSOs.  Symbols are the same as in the upper diagram.\label{midircc} } 
\end{figure}

There were 22 sources detected in all four IRAC channels (with errors $\leq$ 0.1 in [3.6] and [4.5] and errors $\leq$ 0.2 in [5.8] and [8.0]) after bandmerging with the IRSF data and after extragalactic rejection criteria were applied.  They are plotted in the [3.6]$-$[4.5] vs. [5.8]$-$[8.0] diagram in the upper portion of Figure \ref{midircc}.  The loci for main sequence stars (or young stars with little excess emission) and Class II sources are plotted as a \emph{dotted box} and a \emph{dashed box}, respectively (\citealt{Allen:2004p1016}; \citealt{Hartmann:2005p433}).  Both NIR excess sources fall in the Class I region of the MIR color-color diagram.  Their physical locations in the Horsehead are noteworthy: both are embedded in the western bright limb near the cloud/H \small II \normalsize region interface.  B33-1 was previously identified as a young star (\citealt{Reipurth:1984p544}; \citealt{Pound:2003p235}), but its IR class was not ascertained.  Three other sources fall in the Class II region of Figure \ref{midircc}: B33-21, B33-25, and B33-31.   None of the three sources exhibit a NIR excess.  The [3.6]--[4.5] vs. [4.5]--[5.8] color-color diagram is plotted in the lower portion of Figure \ref{midircc}.  The spread in IR classes is much smaller than that of the [3.6]--[4.5] vs. [5.8]--[8.0] color-color diagram, so that plot is not used to identify candidate young stars in this study.  Nevertheless, it is instructive to note that four of the five MIR excess sources appear to have some excess emission in this diagram relative to the rest of the sample.  The symbols are the same as the upper plot and all five MIR excess sources are labeled.

\subsubsection{IRSF and IRAC Combined Color-Color Diagrams}

The combination of NIR and MIR colors has been shown to be useful in identifying bona fide IR excess sources (\citealt{Hartmann:2005p433}; \citealt{Winston:2007p489}).  The 8.0 $\mu$m band is particularly sensitive to PAH emission, which has relevant emission features at 3.3, 6.2, 7.7, and 8.6 $\mu$m (\citealt{Wu:2005p1506}; \citealt{Peeters:2004p300}).  Color-color diagrams without the 8 $\mu$m band are least affected by background extragalactic contaminants that may have been missed in previous selection criteria.  The shorter wavelength IRAC channels are also more sensitive; consequently any young stars with warm disks that did not display a NIR excess and that may not have been detected in the 8 $\mu$m channel should reveal themselves in the combined NIR and MIR color-color diagrams.

\begin{figure}
  \resizebox{\textwidth}{!}{\includegraphics{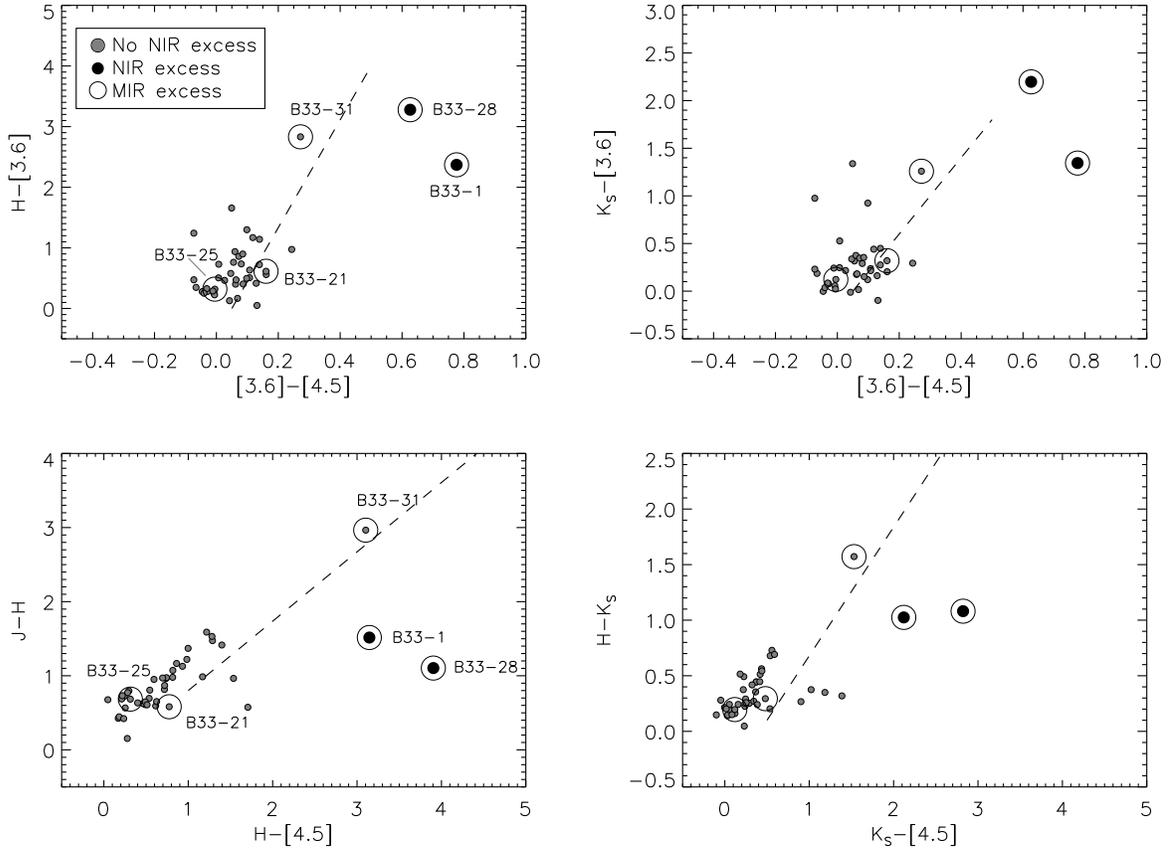}}
  \caption{NIR/MIR color color diagrams showing the positions of NIR excess sources from Figure \ref{jhkcc} (\emph{filled black circles}) and MIR sources from Figure \ref{midircc} (\emph{open circles}).  All sources not exhibiting an IR excess are plotted as \emph{filled gray circles}.   The dashed lines represent the break between normal and IR excess stars (Hartmann et al. 2005; Winston et al. 2007).  The slope of the dashed line is the direction of the reddening vector for each diagram.  \label{nir_mircc1} } 
\end{figure}

\begin{figure}
  \resizebox{\textwidth}{!}{\includegraphics{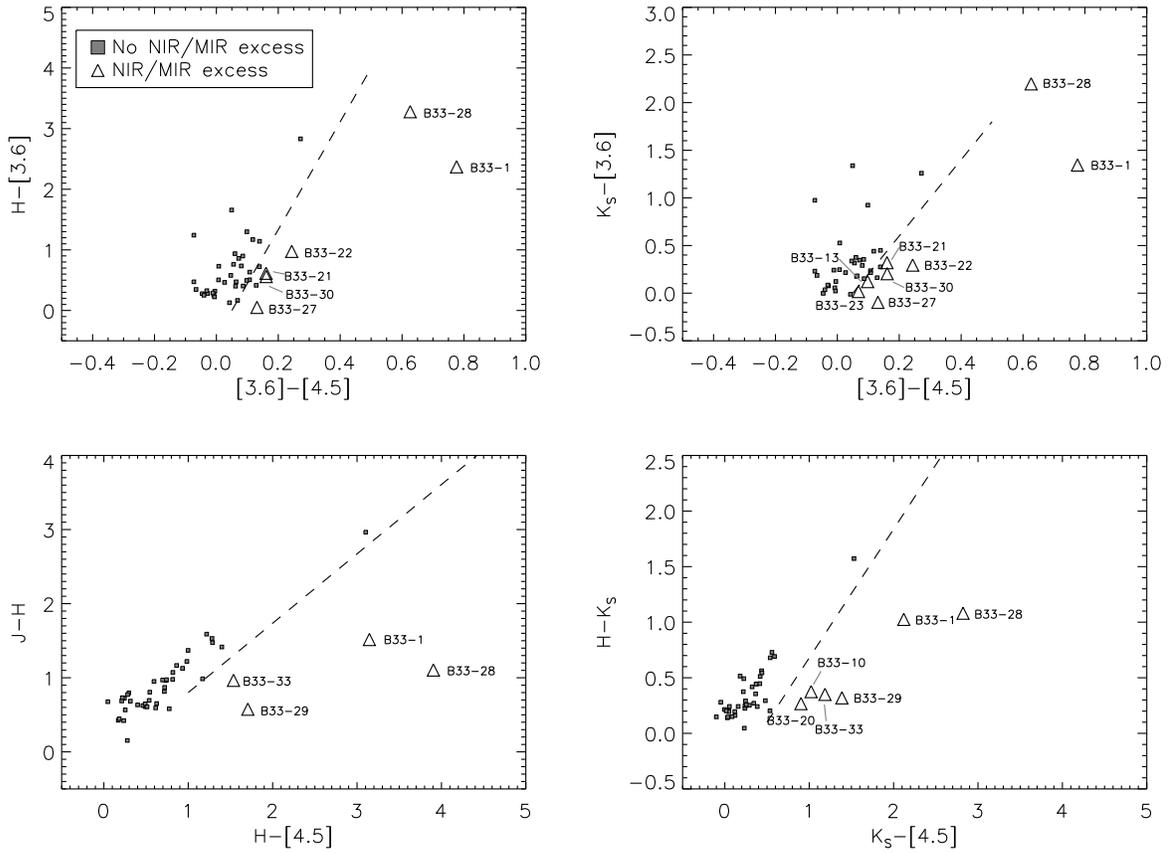}}
  \caption{Similar to Figure \ref{nir_mircc2}, but emphasizing the IR excess sources in each diagram.  Sources exhibiting an IR excess are plotted as \emph{open triangles} and are labeled; all others are represented as \emph{filled squares}. There are 12 unique sources that exhibit an excess in at least one diagram.  \label{nir_mircc2} } 
\end{figure}

We use four combined NIR/MIR color-color diagrams in this analysis:  \emph{H}$-$[3.6] vs. [3.6]$-$[4.5], \emph{K$_S$}$-$[3.6] vs. [3.6]$-$[4.5], \emph{J}$-$\emph{H} vs.  \emph{H}$-$[4.5], and \emph{H}$-$\emph{K$_S$} vs. \emph{K$_S$}$-$[4.5].  The four color-color diagrams are plotted twice for clarity.  In Figure \ref{nir_mircc1} the sources with NIR and MIR excesses are emphasized, whereas Figure \ref{nir_mircc2} labels all sources with excesses in the individual NIR/MIR plots.  \citet{Hartmann:2005p433} define empirical regions of IR excess in the \emph{H}$-$[3.6] vs. [3.6]$-$[4.5] and \emph{K$_S$}$-$[3.6] vs. [3.6]$-$[4.5] diagrams using known young stars in the Taurus star-forming region.  In their work there are clean breaks between Weak-lined T Tauri stars (Class III IR sources) and Classical T Tauri stars (Class II IR sources).  These breaks are represented as \emph{dashed lines} in the top two plots of Figures \ref{nir_mircc1} and \ref{nir_mircc2}; they simultaneously display the slope of the reddening vector for each plot.  We selected sources  with [3.6]$-$[4.5] colors $>$ 1 $\sigma$ to the right of the dashed lines as having IR excesses.  The bottom two plots of Figures \ref{nir_mircc1} and \ref{nir_mircc2} are the \emph{J}$-$\emph{H} vs.  \emph{H}$-$[4.5] and \emph{H}$-$\emph{K$_S$} vs. \emph{K$_S$}$-$[4.5] color-color diagrams.  The 45 sources with NIR and 4.5 $\mu$m detections are plotted.  The \emph{dashed lines} mark the separation of sources with and without IR excesses and also display the slope of the reddening vector for each diagram (from \citealt{Winston:2007p489}).  Sources with \emph{H}$-$[4.5] and \emph{K$_S$}$-$[4.5] colors $>$ 1 $\sigma$ to the right of the dashed lines (for the bottom-left and bottom-right plots, respectively, in Figures \ref{nir_mircc1} and \ref{nir_mircc2}) have IR excesses.  Twelve unique sources exhibit an excess in these diagrams, eight of which show an excess in two or more of the four plots.  The two sources that have combined NIR and MIR  excesses (B33-1 and B33-28) exhibit an excess in all four of the NIR/MIR color-color diagrams.

\subsection{Near-Infrared Color-Magnitude Diagrams}

\begin{figure}
  \resizebox{\textwidth}{!}{\includegraphics{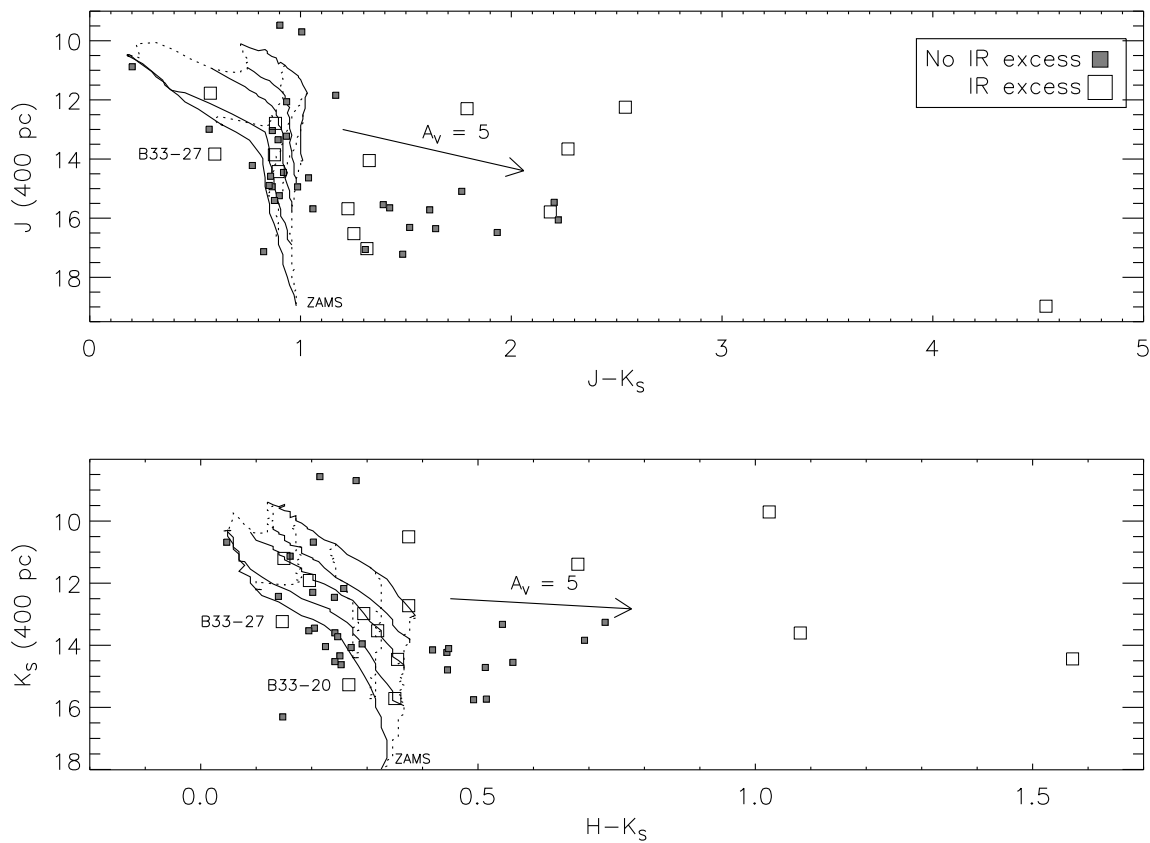}}
  \caption{NIR color-magnitude diagrams showing the positions of the 45 sources with NIR and MIR detections.  The pre-main sequence evolutionary tracks are from Baraffe et al. (1998).  Isochrones are plotted as \emph{solid lines}  for 1, 4, 10, 40,  and 500 Myr (ZAMS) from upper right to lower left.  Iso-mass tracks are plotted as \emph{dotted lines} for 0.08, 0.2, 0.4, 0.8, and 1.4 M$_\odot$ (from lower right to upper left).  The tracks were converted from the CIT photometric system to the 2MASS system using the relations in Carpenter (2001).  They were then shifted to a distance of 400 pc.  Sources with an IR excess in at least one color-color diagram are plotted as \emph{open squares}; those with no excess are plotted as \emph{filled squares}.  The two sources to the left of the ZAMS are inconsistent with being young and at a distance of the Horsehead.  They are therefore rejected as being candidate YSOs.  Many of the IR excess sources show large amounts of reddening.  The reddening vector is derived from Tokunaga (2000). \label{cmd} } 
\end{figure}

In Figure \ref{cmd} we present the $J$ vs. $J$--$K_S$ and $K_S$ vs. $H$--$K_S$ color-magnitude diagrams for the 45 sources with NIR and MIR detections.  The pre-main sequence evolutionary tracks of \citet{Baraffe:1998p160} are overplotted.  We first transformed the evolutionary tracks from the CIT system to the 2MASS system using the relations in \citet{Carpenter:2001p159} and then shifted them to a distance of 400 pc.  Photometry is not corrected for extinction.  We can therefore only identify those stars which cannot be both young and at a distance of 400 pc by selecting against sources to the left of the Zero-Age Main Sequence (ZAMS; left-most isochrone at 500 Myr in  Figure \ref{cmd}).  Two candidate YSOs lie in this region and are consequently rejected as being young.

\subsection{Summary of YSO Identification}

\begin{figure}
  \resizebox{\textwidth}{!}{\includegraphics{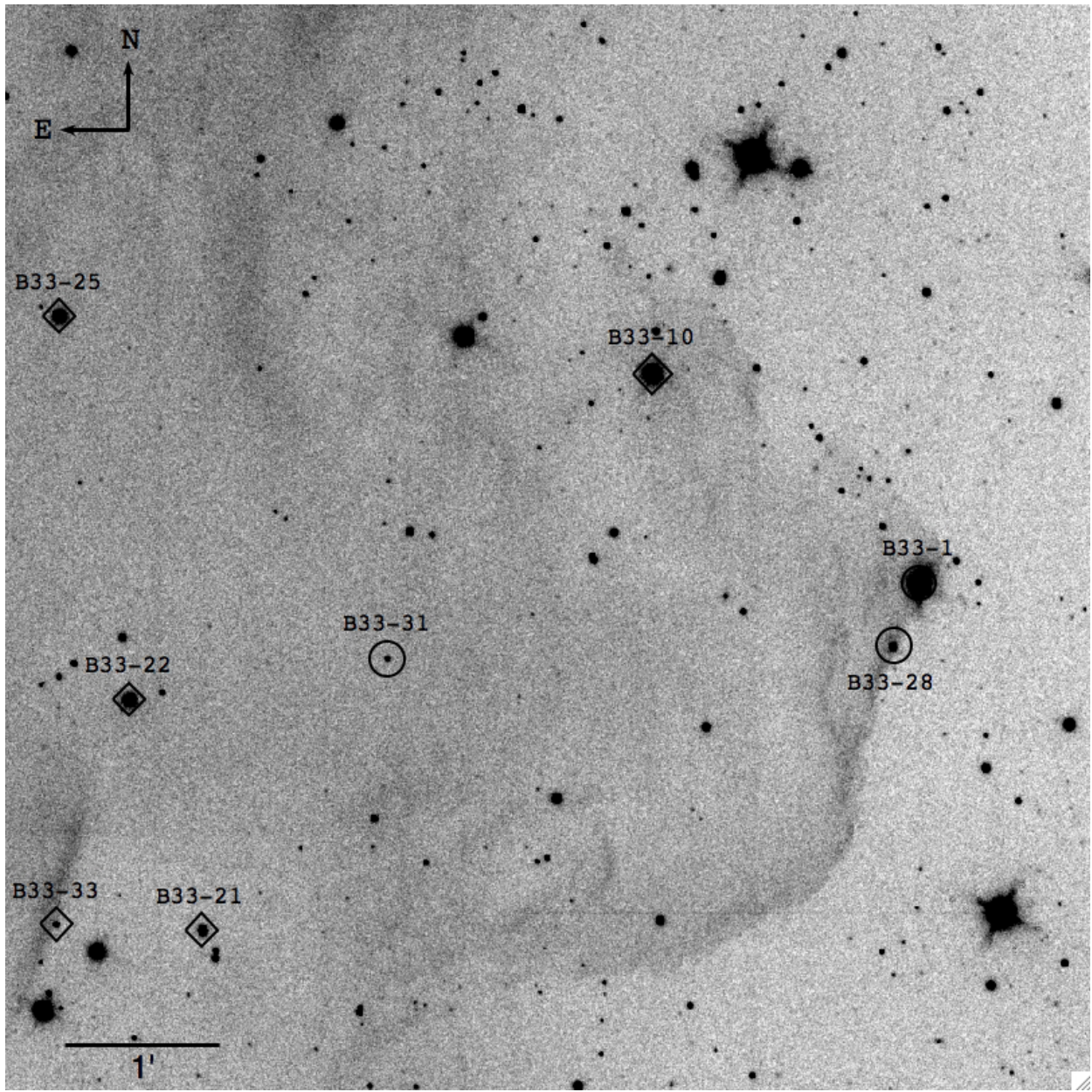}}
  \caption{Locations of bona fide YSOs (\emph{circles}) and candidate YSOs (\emph{diamonds}) overplotted on the $H$-band image of the Horsehead.  The two flat-spectrum sources are located at the western interface of the cloud/H \small II \normalsize region (B33-1 and B33-28).  \label{plotyso} } 
\end{figure}

Sources are considered to be \emph{candidate} YSOs if they have an IR excess in at least two color-color diagrams.  If a source has an excess only in the MIR color-color diagram but has a spectral index indicative of a circumstellar disk ($\alpha_{IRAC}$ $>$ --2.56), then it too is considered a candidate YSO.   Candidate YSOs with a shorter-wavelength color excess must also have an excess at longer wavelengths (see $\S$ 3.1 and Figures \ref{jhkcc}, \ref{midircc}, \ref{nir_mircc1}, \ref{nir_mircc2}).  \emph{Bona fide} YSOs are distinguished from candidate YSOs as having particularly strong IR excess emission and spectral energy distributions (see below) that resemble typical star + disk systems.  Our final requirement  is that candidate young stars must lie to the right of the ZAMS in the NIR color-magnitude diagrams (Figure \ref{cmd}).  A summary of the results is presented in Table \ref{tabsummary}.  Three of the fourteen sources with an excess in at least one color-color diagram are bona fide young stars, two of which (B33-1 and B33-28) have IR excesses and SEDs consistent with flat-spectrum objects in an evolutionary phase between Class I and Class II.  Five of the fourteen are considered candidate young stars.  We continue the nomenclature developed by \citet{Reipurth:1984p544} for IR sources in the Horsehead.  They began with ``B33-1'' and ended at ``B33-26.''  Our new source identifiers begin at ``B33-27'' and run through ``B33-33,'' with count increasing in right ascension.  Only the relevant sources in this study are numbered.  Photometry for the eight bona fide and candidate YSOs is presented in Table 2 and their positions are overplotted on the $H$-band image of the Horsehead in Figure \ref{plotyso}.

\begin{figure}
  \resizebox{\textwidth}{!}{\includegraphics{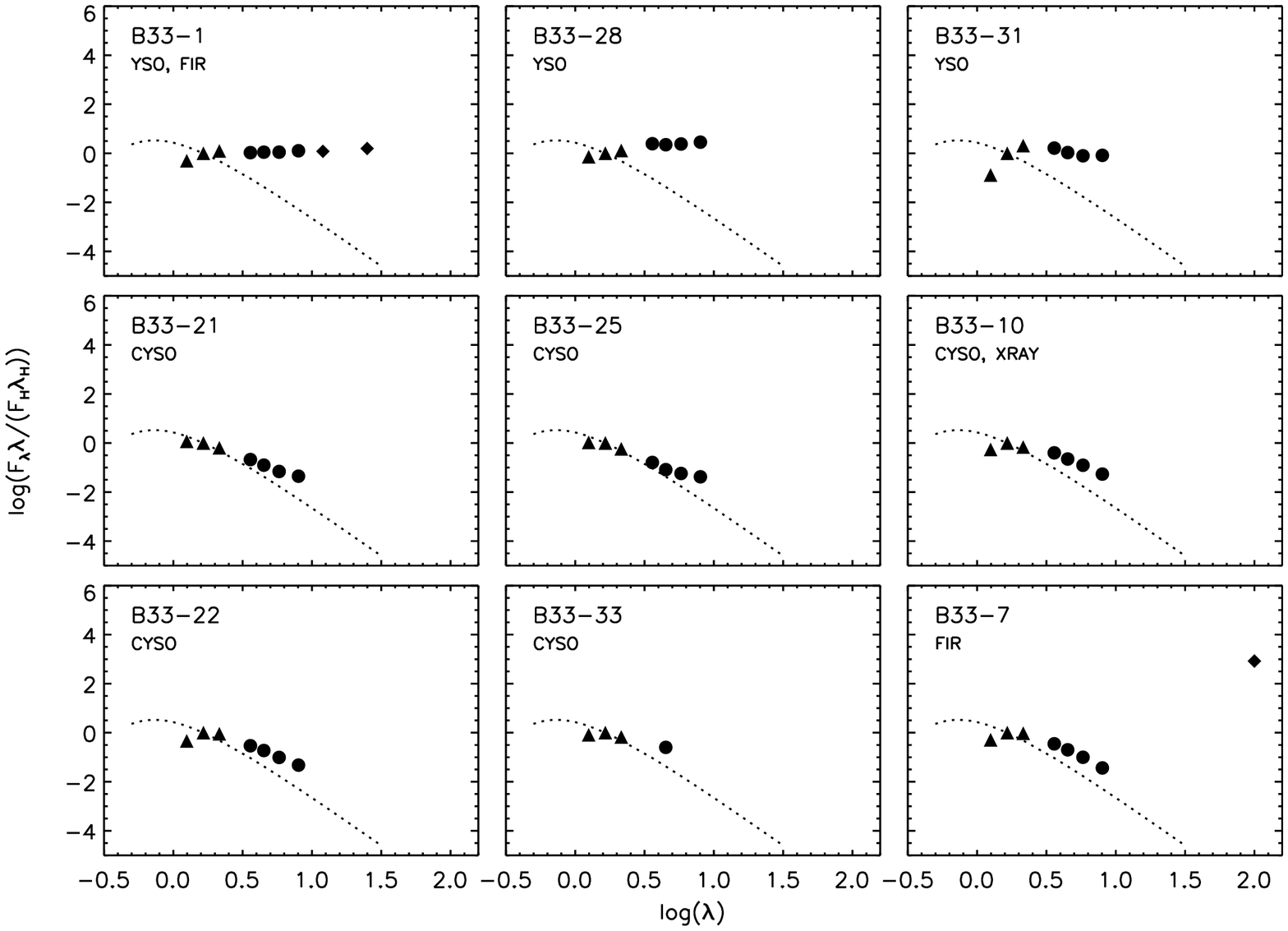}}
  \caption{SEDs of bona fide YSOs (``YSO''), candidate YSOs (``CYSO''), X-ray sources (``XRAY''), and/or far-infrared sources (``FIR'').  IRSF, IRAC, and IRAS data are labelled as \emph{filled triangles}, \emph{filled circles}, and \emph{filled diamonds}, respectively.  Magnitudes are converted to fluxes using the absolute calibration of \citet{Rieke:2008p11100}, who modified previous calibrations from \citet{Cohen:2003p210} for 2MASS and \citet{Reach:2005p11118} for IRAC.  The zero-point offsets used were  1617, 1073, 675, 280.9, 179.7, 115.0, and 65.09 Jy for the 1.25, 1.65, 2.15, 3.6, 4.5, 5.8, and 8.0 $\mu$m bands. SEDs were normalized to the $H$-band.  \emph{Dotted lines} display a 4000 K blackbody, also normalized to the $H$-band data point.  Note the odd 100 $\mu$m point of B33-7 (which does not show an excess in any color-color diagram); this point is probably caused by coincident IR cirrus emission.   The IRAS data for this source was flagged as having a strong likelihood of contamination.  \label{sed} } 
\end{figure}

Spectral energy distributions (SED) for the bona fide and candidate young stars are plotted in Figure \ref{sed}.  The SED of one source that did not have an IR excess but that may be the IR counterpart to IRAS 05384--0229 is plotted in the same figure (B33-7; see $\S$ 3.4.1 for details).  The conversion between magnitudes and flux densities was done using the new absolute calibration of the 2MASS system and \emph{Spitzer}/IRAC magnitude system by \citet{Rieke:2008p11100}, updated from the popular \citet{Cohen:2003p210} and \citet{Reach:2005p11118} calibrations for 2MASS and IRAC, respectively.  All SEDs are normalized to the $H$-band flux, which is less affected by reddening than the $J$-band.  A blackbody spectrum with a temperature of 4000 K and scaled to the $H$-band flux is overplotted as a reference (\emph{dotted lines}).  

The  evolutionary stage of the star plus disk/envelope system defines the overall shape of an SED and consequently the IR class (\citealt{Adams:1987p77}; \citealt{Lada:2006p519}).  A common MIR classification system is based on the value of the 3.6 to 8.0 $\mu$m spectral index, defined as $\alpha =  d\log{\lambda F_{\lambda}}/d\log{\lambda}$.  In this study we generally use the following classification scheme, although we also take into account the positions in color-color diagrams to assign an IR class (Table \ref{tabsummary}): IRAC Class I sources have $\alpha_{IRAC}$ $>$ 0.3, IRAC flat-spectrum sources have $-$0.3 $\leq$ $\alpha_{IRAC}$ $<$ 0.3, IRAC Class II sources have $-$1.8 $\le \alpha_{IRAC} <$ $-$0.3, sources with transition (``anemic'') disks have $-$2.56 $\leq$ $\alpha_{IRAC}$ $<$ $-$1.8, and IRAC Class III sources have $\alpha_{IRAC} < $ $-$2.56 (\citealt{Lada:2006p519}).  Spectral indices are calculated by fitting a line to the four IRAC fluxes in log($\lambda$$F_{\lambda}$) vs. log($\lambda$) space.  The slope of the line is the spectral index.  Values of the spectral indices for candidate young stars are listed in Table \ref{tabsummary}.  Note that an M-dwarf photosphere has a spectral index of --2.66 (\citealt{Lada:2006p519}).

\subsubsection{Counterparts to IRAS and ASCA Sources}

Protostars have warm disk/envelope systems that reprocess stellar light into longer wavelength radiation.  They are therefore often identified in FIR surveys of star-forming regions.  More evolved young stars both with and without disks exhibit X-ray emission likely caused by intense magnetic activity generated by a strong dynamo effect.  Large flaring events are thought to be the origin of the typical X-ray variability of Classical T Tauri stars.  The detection of both FIR IRAS sources and ASCA X-ray sources in the vicinity of the Horsehead Nebula supports the notion that there is ongoing star formation in this region.  It has been suspected that the IRAS and ASCA sources are young stars, but large positional uncertainties in those data have prevented the identification of most of their optical/IR counterparts.  Three IRAS sources were detected in the Horsehead: IRAS 05383--0228, IRAS 05384--0229, and IRAS 05386--0229 (photometry is given in \citealt{Pound:2003p235}).  B33-1 has been identified as the optical/IR counterpart of IRAS 05383--0228.  IRAS 05384--0229 was only detected in the 100 $\mu$m band and likely originates from infrared cirrus: all three sources have contamination flags that indicate a high probability of being caused by cirrus emission.  These contamination flags are based on a large number of nearby 100 $\mu$m-only sources, the presence of large-scale structure in the 100 $\mu$m emission map, and a high surface brightness at 100 $\mu$m in the region.  IRAS 05386--0229 was only detected in the 12 and 25 $\mu$m bands of IRAS.  It too has high cirrus contamination flags.  Nevertheless, we will compare the positions of the three IRAS sources with the locations of the candidate young stars identified in this work to search for possible YSO counterparts.  Two X-ray sources in this region were detected by ASCA by two different groups: [NY99] C-18 and [NY99] C-20 by \citet{Nakano:1999p502}, and [YKK2000] A4 by \citet{Yamauchi:2000p225}.  [NY99] C-18 was detected by the Gas Imaging Spectrometer (GIS) and the Solid-State Imaging Spectrometer onboard ASCA.  [NY99] C-20 was detected by the GIS only, as was [YKK2000] A4.  It is likely that [NY99] C-18 and [YKK2000] A4 are the same source as their separation is only 18$\farcs$5 while their positional uncertainties are both about 1\arcmin.  \citet{Yamauchi:2000p225} suggest B33-3, B33-4, B33-5, B33-9, and B33-10 as possible counterparts to [YKK2000] A4.

\begin{figure}
\begin{center}
  \resizebox{.75\textwidth}{!}{\includegraphics{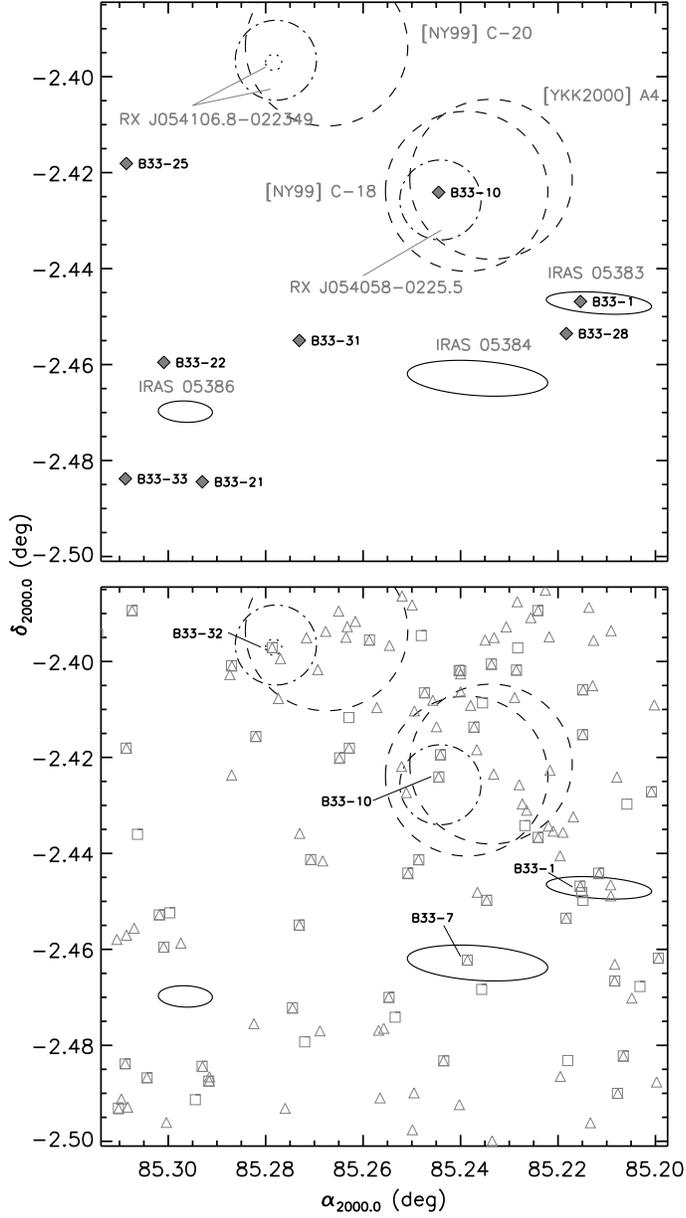}}
  \caption{Identification of X-ray and far-IR counterparts.  In the upper panel we plot the error circles (ellipses for IRAS sources) for the two X-ray sources detected by both ASCA and ROSAT (from archival observations) together with the three IRAS sources in the field.  Overplotted as \emph{filled diamonds} are the bona fide and candidate young stars identified in this survey.   In the lower panel all IRSF sources with color errors $<$ 0.15 mag (\emph{open triangles}) and IRAC sources with errors $<$ 0.1 mag in the first two bands and errors $<$ 0.2 mag in the last two bands (\emph{open squares}) are overplotted in gray.  We identify possible IR counterparts to both X-ray sources and to two IRAS sources.  However, one IRAS source for which we identify a counterpart (IRAS 05384--0229; B33-7) was only detected in the 100 $\mu$m band and has a high probability flag for being cirrus emission.  The abnormal SED of B33-7 with the IRAS 100 $\mu$m flux supports the idea that this is likely caused by IR cirrus.  \label{plotyso_both} } 
 \end{center}
\end{figure}

In Figure \ref{plotyso_both} we plot the positional uncertainties of the IRAS and ASCA observations.  The upper plot shows the positions of the bona fide and candidate YSOs, while the lower plot displays all the NIR and MIR source detections.   In the lower plot, IRSF and IRAC detections \emph{before merging the two data sets} and \emph{before removing extragalactic contaminants} are displayed.  We required simply that the IRAC sources have errors $\leq$ 0.1 in channels 1 and 2 and errors $\leq$ 0.2 in channels 3 and 4.  The IRSF sources with color errors $\le$ 0.15 are plotted.  We used a 1$\arcmin$  error circle for the ASCA observations (\citealt{Yamauchi:2000p225}).  We searched the archival ROSAT database and found that two pointed observations were made centered near the Horsehead with the Position Sensitive Proportional Counter (PSPC) and the High Resolution Imager (HRI) detectors.  The observation sequence for the PSPC frame is RP900189 and for the HRI frame is RH201148 (PI: J.H.M.M. Schmitt); observations were made between 19 and 20 Sept 1991 and between 12 and 14 Sept 1992 for the PSPC and HRI frames, respectively.  Two X-ray sources are detected nearly on-axis in the PSPC observation (RX J054058--0225.5 and RX J054106.8--022349) and only one was detected with the HRI (RX J054106.8--022349).  We use the IAU naming convention for both sources.  This non-detection of RX J054058--0225.5 in the HRI observation suggests that it is variable in X-rays.  We determined the centroid of each source and plotted their positional uncertainties in Figure \ref{plotyso_both}.  The centroids are given in Table \ref{counterpart}.  We used error circles of 30$\arcsec$ for the PSPC frame (\citealt{Barbera:2002p512}; \citealt{Bocchino:2001p513}) and 6$\arcsec$ for the HRI frame (\citealt{Bocchino:2001p513}).

We invoke Ockham's razor and assume that the [YKK2000] A4, the NY[99] C-18, and RXJ054058--0225.5 are the same source, as are [NY99] C-20 and RXJ054106.8--022349.  We identify B33-10 as the likely counterpart of the RXJ054058--02225.5 group of X-ray sources (see Figure \ref{plotyso_both}, bottom), which sits neatly in all three error circles.  Similarly, we identify B33-32 as the likely counterpart of the other group of X-ray sources centered around RXJ054106.8--022349.   It is the nearest IR source to the centroid of RXJ054106.8--022349 out of the two IR sources detected by both IRSF and IRAC within the RXJ054106.8--022349 error circle, the other source being B33-9 (not labeled).  B33-1 is re-identified as the likely counterpart to IRAS 05383--0228.  No candidate YSOs are located in the error ellipses of the other two IRAS sources, but B33-7 lies near the centroid of the IRAS 05384--0229 detection.  While this IRAS source is likely caused by cirrus emission and is probably not a real point source, we simply note that B33-7 lies near the center of the IRAS error ellipse.  No counterparts are identified for IRAS 05386--0229.  These results are tabulated in Table \ref{counterpart}.

\section{Discussion}

We identify two bona fide YSOs (B33-1 and B33-28) at the western limb of the Horsehead Nebula as flat-spectrum protostars based on their positions in color-color diagrams, their spectral indices, and the overall shape of their SEDs.  That their SEDs are so similar argues that they are in the same evolutionary phase, and consequently that they formed at nearly the same time.  Their apparent separation is 26$\arcsec$, or $\sim$ 10400 AU (0.05 pc) at 400 pc.  B33-1 appears to be emerging from an optically-visible cavity (see Figure 5 in \citealt{Pound:2003p235}).  Immediately west of the cavity is a small filamentary structure that is barely detected in our NIR images.  It is possible that this thin filament is an an outflow from B33-1.  B33-28 is optically-invisible and so is either embedded or is emerging on the other side of this nearly edge-on pillar.  

\begin{figure}
\begin{center}
  \resizebox{.8\textwidth}{!}{\includegraphics{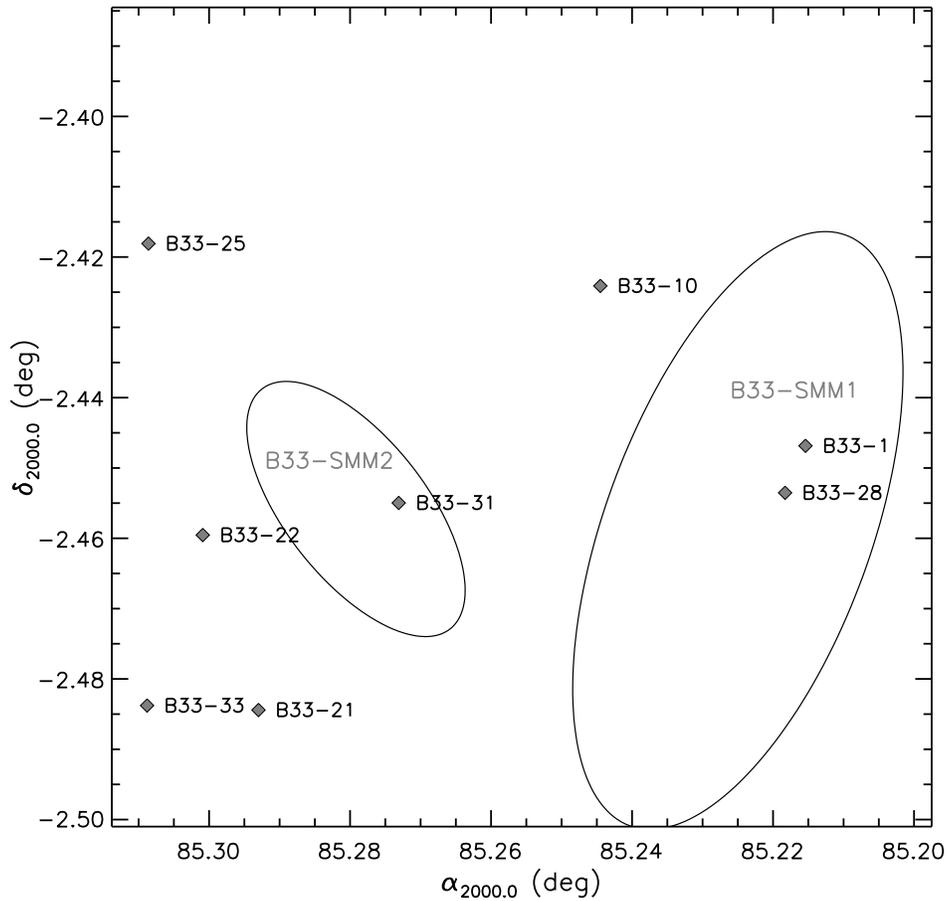}}
  \caption{A comparison between the two sub-mm clumps identified by Ward-Thompson et al. (2006) and the bona fide and candidate young stars identified in this survey (\emph{filled diamonds}).  The \emph{solid lines} show the best-fitting ellipses to the 3 $\sigma$ sub-mm contours of the B33-SMM1 and B33-SMM2 clumps.  B33-1 and B33-28 are near the periphery of the centroid of B33-SMM1, as is B33-31 for B33-SMM2. We expect sources that may be the products of star formation in these clumps to be in near the clumps' centroids; we do not identify any young stars that are associated with the clumps.  \label{plotsmm} } 
  \end{center}
\end{figure}

\citet{WardThompson:2006p540} recently presented observations of the Horsehead in 450 $\mu$m and 850 $\mu$m in which they discussed the stability of two dense condensations found in the neck and the head regions (B33-SMM1 and B33-SMM2).  They calculate the virial equilibrium of both clumps and conclude that B33-SMM1 has been affected by the ionizing radiation of $\sigma$ Ori and may undergo triggered star formation, while B33-SMM2 is likely a pre-existing clump in gravitational equilibrium and may be a protostellar core.  In Figure \ref{plotsmm} we overplot the locations of both clumps along with the bona fide and candidate young stars from this study to search for any spatial coincidence.  The best-fitting ellipses that encompass the 3 $\sigma$ sub-mm contours are drawn as \emph{solid lines} (parameters are given in \citealt{WardThompson:2006p540}).   B33-1 and B33-28 are located at the periphery of the B33-SMM1 condensation.  Likewise, B33-31 is not centered on the B33-SMM2 clump, although it does border that region.  B33-31 is apparently an extremely embedded source given its highly reddened positions in all of the color-color diagrams.  We estimate a reddening of $A_V$ $\sim$ 22 mag from its location in the NIR color-color diagram down to the locus of Classical T Tauri stars.  That the centroids of B33-31 and B33-SMM2 do not coincide bolsters the notion that B33-SMM2 is a pre-stellar core that has yet to undergo star formation.

The overall picture of star formation in the vicinity of the Horsehead is not straightforward.  It appears that several generations of young stars occupy the line-of-sight path to this pillar, but it remains unclear which sources are actually \emph{associated} with the Horsehead (besides the protostars).  One way to better understand the spatial proximity of candidate YSOs to the Horsehead is to look at their reddening in the NIR color-color diagram.  If one assumes that B33-21 and B33-25 should lie near the Classical T Tauri star locus in Figure \ref{jhkcc}, then both sources appear to be in the foreground of the cloud as there is virtually no indication of reddening.  The X-ray source B33-10 is slightly reddened ($A_V$ $\sim$ 4 mag) and may be partially embedded or may be a background source seen through tenuous material at the ``nose'' of the Horsehead.  In the latter scenario it is possible that B33-10 is a part of the extended $\sigma$ Ori cluster, as members have been found almost all the way to the Horsehead itself (\citealt{Hernandez:2007p387}).  In fact, it could be argued that the newest members of the $\sigma$ Ori cluster are just emerging at the tip of the Horsehead.  This becomes somewhat arbitrary though as all these disparate populations eventually blend  into each other.  The other X-ray source B33-32 sits directly on the mid-M portion of the main-sequence locus in Figure \ref{jhkcc}.  This raises the possibility that B33-32 might be a foreground M-dwarf, as magnetic M-dwarfs are strong X-ray emitters.  Follow-up spectroscopic observations are required to fully understand most of the young stellar population in this region.

Unfortunately, this study sheds little light on the origin of the unique structure of the Horsehead.  Nevertheless, we are able to state that there is no evidence of an IR excess in the previously suspected young star B33-6, which it was suggested had formed the ``jaw'' cavity through an outflow (\citealt{Reipurth:1984p544}).  Neither do we find evidence for small-scale sequential star formation within $\sim$ 1$\farcm$5 (0.17 pc) of the tip of the Horsehead, although we do find \emph{simultaneous} star formation at its leading edge.  It is probably not possible to conclusively state that the Horsehead is a site of triggered star formation as there are very few non-degenerate predictions of spontaneous and triggered star formation theories.  Nevertheless, the observation that most relatively massive irradiated elephant trunks \emph{only show young stars at their tips} suggests that what we are observing is indeed the result of a triggering process.  Noting that the sum-mm clump B33-SMM1 was probably formed from the direct influence of $\sigma$ Ori (\citealt{WardThompson:2006p540}) and that the two simultaneously-formed protostars positionally coincide with the leading edge of this pillar suggests that these protostars were probably formed as a result of the influence of $\sigma$ Ori.  The formation timescale of the Horsehead is $\sim$ 10$^5$ yrs (\citealt{Pound:2003p235}) which is similar to the estimated ages of flat-spectrum protostars (\citealt{Kenyon:1990p13275}; \citealt{Greene:1995p13218}).  These similar timescales are consistent with the picture of B33-1 and B33-28 forming when the pillar itself began to take shape.  Our observations closely resemble the simulations of \citet{Gritschneder:2007p1226}, who show that ionizing radiation can lead to the formation of pillars similar in size to the Horsehead, and that core formation and collapse at the tips of such pillars can occur.

 \section{Conclusion}
 
 We present near-infrared (IRSF/SIRIUS) and mid-infrared (Spitzer/IRAC) observations of the Horsehead Nebula in order to study the state and extent of star formation in this nearby pillar.  We use six near- to mid-infrared color-color diagrams to select candidate young stars based on their infrared excesses.  We also ensure that their positions in near-infrared color-magnitude diagrams are consistent with youth.  In total we find two flat-spectrum protostars (B33-1, B33-28), a Class II YSO, and five candidate young stars.  Spectral energy distributions indicate thick disk/envelope systems for the protostars and only modest excesses in most of the candidate young stars.  We identify the infrared counterparts of two X-ray sources in the region and the counterpart to one of three IRAS sources, the other two likely being caused by IR cirrus emission.  
 
 Star formation in the Horsehead is ongoing and is likely triggered by the influence of the nearby $\sigma$ Orionis OB system.  We find no evidence for sequential star formation in this region, but the two protostars are in the same evolutionary phase and were probably formed from the collapse of a core that has since been photoevaporated away.  The formation timescale of the Horsehead is also similar to the estimated ages of protostars.  These observations are consistent with the Radiation-Driven Implosion model of triggered star formation, although it is not possible to tell whether star formation in this region is simply an accelerated version of what would have occurred naturally, or whether star formation is taking place where it otherwise not have spontaneously done so.  However, one available clue is the observation of an externally-compressed sub-mm clump at the western limb of the Horsehead (B33-SMM1), whose properties are discussed by \citet{WardThompson:2006p540}.  This clump is out of virial equilibrium and, as such, may have been compressed by the influence of $\sigma$ Ori.  That such an influence is evident immediately behind the locations of these protostars may suggest that they were formed in a similar fashion.  
 
 The candidate young stars surrounding the Horsehead range from foreground to partly embedded to fully embedded objects.  Their association with the Horsehead remains unclear.  Follow-up spectroscopy and longer-wavelength photometry will be useful to confirm their status as young stars.

\acknowledgments
We thank the referee for the useful feedback and Bo Reipurth for his helpful comments on a draft of this manuscript.  This work makes use of data products from the Two Micron All Sky Survey, which is a joint project of the University of Massachusetts and the Infrared Processing and Analysis Center/California Institute of Technology, funded by the National Aeronautics and Space Administration and the National Science Foundation.  This research made use of the SIMBAD database, operated at CDS, Strasbourg, France.  BPB extends special thanks to the Massachusetts Space Grant Consortium and the Tufts Summer Scholars Program for their support.  BPB also wishes to thank Bill Oliver and Robert Willson of Tufts University for their advice.  

{\it Facilities:} \facility{Spitzer (IRAC), IRSF (SIRIUS)}

\bibliography{dec08}
\bibliographystyle{apj}

\clearpage

\clearpage

\clearpage

\begin{deluxetable}{lccccccccccccc}
\tabletypesize{\scriptsize}
\tablewidth{0pt}
\tablecolumns{13}
\tablecaption{Summary of Color-Color and Color-Magnitude Diagrams \label{tabsummary}}
\tablehead{
\colhead{}      & \colhead{$\alpha_{J2000.0}$}    &  \colhead{$\delta_{J2000.0}$}    &  \colhead{}         &\colhead{}                 &\colhead{}           &\colhead{}         &\colhead{}          &\colhead{}          &\colhead{}           &\colhead{}          &    \colhead{}                              & \colhead{YSO}  & \colhead{IR}  \\
\colhead{ID} & \colhead{(h m s)}                          &      \colhead{($\circ$ ' '')}                  & \colhead{CC1} &    \colhead{CC2}     & \colhead{CC3} & \colhead{CC4} & \colhead{CC5} & \colhead{CC6} & \colhead{CM1} & \colhead{CM2} & \colhead{$\alpha_{IRAC}$}  & \colhead{Status}     & \colhead{Class}    
}
\startdata

B33-27                               &   05 40 48.20  &  --02 25  37.8     &  N  &   N              & Y     & Y     &  N      &  N    &  N  &  N &   --2.65         & N      & \nodata  \\
B33-1\tablenotemark{a}  &   05 40 51.70  &  --02 26  48.7     &  Y  &   Y              & Y     & Y     &  Y      &  Y    &  Y  &  Y  &      0.20         & YSO   & FS  \\
B33-28                               &   05 40 52.40  &  --02 27  12.7     &  Y   &   Y              & Y     & Y      &  Y     &  Y    &  Y  &  Y  &      0.21         & YSO    & FS \\
B33-29                               &   05 40 57.60  &  --02 24  09.3     &  N   &  \nodata   & N     & N     &  Y     &  Y     &  Y  &  Y  &   \nodata      & N   & \nodata \\
B33-10\tablenotemark{a}                               &   05 40 58.68  &  --02 25  26.8     &  N   &   N             & N     & N     &  N     &  Y     &  Y  &  Y  &  --2.48         & CYSO  & III   \\
B33-13\tablenotemark{a}                                &   05 41 01.14  &  --02 28  12.0    &  N   &   \nodata  & N     & Y     &  N     &  N     &  Y  &  Y  &  \nodata      & N           &  \nodata  \\
B33-20\tablenotemark{a}                                &   05 41 09.97  &  --02 29  11.6    &  N   &    \nodata  & N     & N     &  N     &  Y    &  Y  &  N  &    \nodata    & N          & \nodata   \\
B33-30                                &   05 41 02.09  &  --02 23  44.1    &  N   &   \nodata   & Y     & Y     &  N     &  N     &  Y  &  Y & \nodata        & N   & \nodata  \\
B33-31                               &   05 41 05.53  &  --02 27  17.9     &  N   &    Y             & N     & N     &  N     &  N    &  Y  &  Y  &   --0.84       & YSO    & II   \\
B33-21\tablenotemark{a} &   05 41 10.31  &  --02 29  03.9   &  N   &    Y             & Y     & Y      &  N     &  N    &  Y  &  Y  & --1.97         & CYSO     & II  \\
B33-22\tablenotemark{a} &   05 41 12.23  &  --02 27  34.3   &  N   &    N             & Y     & Y      &  N     &  N    &  Y  &  Y  & --2.31          & CYSO    &  TD  \\
B33-23\tablenotemark{a}                                &   05 41 13.04  &  --02 29  12.2    &  N   &    N             & N     & Y      &  N     &  N   &  Y  &  Y   &    --2.80        & N    &  \nodata          \\
B33-25\tablenotemark{a}                                &   05 41 14.06  &  --02 25  05.1    &  N   &    Y             & N     & N      &  N     &  N   &  Y  &  Y  &     --1.62         & CYSO     & TD \\
B33-33                                 &   05 41 14.12  &  --02 29  01.6   &  N   &    \nodata  & \nodata   & \nodata      &  Y   &  Y  &  Y  &  Y & \nodata   &  CYSO & \nodata   \\

\enddata
\tablecomments{All sources with an IR excess in at least one color-color diagram are listed.  The color-color diagram numbering system is as follows: CC1=$J$--$H$ vs. $H$-$K_S$, CC2=[3.6]--[4.5] vs. [5.8]--[8.0], CC3=$H$--[3.6] vs. [3.6]--[4.5],  CC4=$K_S$--[3.6] vs. [3.6]--[4.5], CC5=$J-H$ vs. $H$--[4.5], and CC6=$H-K_S$ vs. $K_S$--[4.5].  A ``Y'' for the color-color diagrams indicates that the source exhibits an excess in that plot.  Color-magnitude diagrams are as follows: CM1=$J$ vs. $J$--$K_S$, CM2=$K_S$ vs. $H$--$K_S$.  A ``Y'' in the color-magnitude columns means that the position of that source is consistent with being young and at a distance of the Horsehead (i.e. to the right of the ZAMS).  Spectral indices in Column 12 are simply the slope of the best-fitting line to the 4-channel IRAC data in the log($\lambda$F$_{\lambda}$) vs. log($\lambda$) space.  Column 13 indicates the status of the sources as YSOs.  Bona fide YSOs are labelled as ``YSO'' while candidate YSOs have a ``CYSO'' label.  Sources that are not YSOs are labeled ``N''.  The infrared class is listed in Column 14 for flat-spectrum sources (``FS''), Class II sources (``II''), transition disk objects (``TD''), and Class III sources (``III''), which were determined using a combination of SEDs, positions in color-color diagrams, and $\alpha_\mathrm{IRAC}$ indices.)
}
\tablenotetext{a}{Source numbers from Reipurth $\&$ Bally (1984)}

\end{deluxetable}

\clearpage

\begin{deluxetable}{lcccccccc}
\tabletypesize{\scriptsize}
\tablewidth{0pt}
\tablecolumns{9}
\tablecaption{Infrared Photometry \label{phot}}
\tablehead{
\colhead{ID} & \colhead{$J$ ($\sigma_J$)} &     \colhead{$J-H$ ($\sigma_{J-H}$)} &    \colhead{$K_S$ ($\sigma_{K_S}$)} & \colhead{$H-K_S$ ($\sigma_{H-K_S}$)} & \colhead{[3.6] ($\sigma_{3.6}$)} & \colhead{[4.5] ($\sigma_{4.5}$)} & \colhead{[5.8] ($\sigma_{5.8}$)} & \colhead{[8.0] ($\sigma_{8.0}$)} 
}
\startdata

\cutinhead{Bona Fide YSOs}
B33-1\tablenotemark{a}    & 12.25 (0.04)  &  1.52 (0.05) & 9.71 (0.03)     &  1.03 (0.04) &     8.362 (0.002) &    7.586 (0.002) &    6.824 (0.002)  &    5.717 (0.002)  \\
B33-28                                 & 15.79 (0.04)  &  1.10 (0.05)  &13.53 (0.06)    &   1.08 (0.03) &    11.407 (0.018) &   10.781 (0.007) &    9.947 (0.056)  &    8.792 (0.098)  \\
B33-31                                 &  18.98 (0.14) &   2.97 (0.15) & 14.36 (0.08)    & 1.57 (0.05)  &   13.180 (0.010)  &  12.908 (0.009)  &   12.475 (0.076)   &  11.464 (0.159)   \\

\cutinhead{Candidate YSOs}

B33-10                                 &  12.30 (0.04)  & 1.42 (0.05)   &  10.52 (0.02)   &  0.38 (0.03) &     9.583 (0.002)  &    9.484 (0.002)  &    9.358 (0.005)   &     9.294 (0.029)   \\
B33-21                                 & 13.85 (0.02)  &   0.58 (0.04) &12.98 (0.02)  &  0.29 (0.03) &    12.659 (0.005) &   12.499 (0.005) &    12.385 (0.031) &    11.898 (0.073) \\
B33-22                                 & 13.66 (0.04)  &    1.59 (0.06) &11.37 (0.04) &  0.68 (0.02) &    11.096 (0.005) &   10.852 (0.004) &    10.799 (0.036) &    10.616 (0.153) \\
B33-25                                 &  12.80 (0.02)  &   0.69 (0.04)  & 11.92 (0.02)  &  0.20 (0.04) & 11.791 (0.007)  &  11.796 (0.006)  &  11.425 (0.064)    &  10.798 (0.152)   \\
B33-33                                 & 17.03 (0.05)  &  0.97 (0.06)   & 15.71 (0.02)   &   0.35 (0.06) &         \nodata  &   14.528 (0.041) &      \nodata     &   \nodata       \\

\cutinhead{X-ray Counterparts\tablenotemark{b}}
B33-32                                 & 13.04 (0.02)  &   0.61 (0.04)  &12.17 (0.02) &   0.26 (0.03) &   11.925 (0.007) &  11.918 (0.006) & 11.808  (0.068)  &  \nodata \\
\enddata


\tablenotetext{a}{Saturated in the NIR images.  $JHK_S$ photometry is from 2MASS.}

\tablenotetext{b}{B33-10 and B33-32 are likely the IR counterparts to X-ray sources.  Photometry for B33-10 can be found under the ``Candidate YSOs'' section of this table.}

\end{deluxetable}

\clearpage

\begin{deluxetable}{lcccccc}
\tabletypesize{\scriptsize}
\tablewidth{0pt}
\tablecolumns{7}
\tablecaption{Likely Counterparts to Far-IR/X-Ray Sources in the Horsehead \label{counterpart}}
\tablehead{

\colhead{IR}                  & \colhead{$\alpha_{J2000.0}$}   &   \colhead{$\delta_{J2000.0}$} & \colhead{}                             & \colhead{YSO}      & \colhead{}                & \colhead{Type of}    \\
\colhead{Counterpart} & \colhead{(h m s)}                          &      \colhead{($\circ$ ' '')}             & \colhead{$\alpha_IRAC$} & \colhead{Status}   & \colhead{Source ID} & \colhead{Source}

}
\startdata
B33-1\tablenotemark{a}        &  05 40 48.20  &  --02 25  37.8  & 0.20          &   YSO & IRAS 05383--0228         & FIR \\
B33-7?\tablenotemark{a,b}  &  05 40 57.26  & --02 27 44.0    & --2.84       & N         & IRAS 05384--0229         & FIR    \\
B33-10\tablenotemark{a}     &  05 40 58.68  & --02 25 26.7   &   --2.48     & CYSO   & RX J054058--0225.5     & X     \\
                                                   &                        &                          &                   &               &     [NY99] C-18                 & X       \\
                                                  &                         &                          &                   &               & [YKK2000] A4                  & X      \\
B33-32                                     & 05 41 06.89   &  -02 23 49.6    &  \nodata  & N           & RXJ054106.8--022349  & X      \\
                                                  &                         &                           &                  &               & [NY99] C-20                     & X     \\
\enddata
\tablecomments{Probable infrared counterparts to far-infrared (FIR) and X-ray sources.  Labels for YSO status are the same as for Table \ref{tabsummary}.}

\tablenotetext{a}{Source numbers from Reipurth $\&$ Bally (1984)}
\tablenotetext{b}{B33-7 lies near the centroid of IRAS 05384--0229, but that IRAS source is probably caused by cirrus emission.  See $\S$ 3.4.1 for details}
\end{deluxetable}

\end{document}